\documentclass[%
 reprint,
 amsmath,amssymb,
 aps,
]{revtex4-2}

\usepackage{graphicx}
\usepackage{dcolumn}
\usepackage{bm}
\usepackage{hyperref}

\usepackage{enumitem}
\usepackage{amsthm}
\newtheorem{theorem}{Theorem}

\newtheorem{definition}{Definition}

\begin{document}

\preprint{APS/123-QED}

\title{Symmetries of many-body systems imply distance-dependent potentials}

\author{Jonathan Utterson}
\email{utterson@maths.ox.ac.uk}
\author{Radek Erban}%
\email{erban@maths.ox.ac.uk}
\affiliation{%
 Mathematical Institute, University of Oxford, Radcliffe Observatory Quarter, Woodstock Road, Oxford OX2 6GG, United Kingdom
}%

\date{\today}

\begin{abstract}
\noindent Considering interatomic potential $U({\mathbf q})$ where ${\mathbf q} = [{\mathbf q}_1, {\mathbf q}_2, \dots, {\mathbf q}_N] \in {\mathbb R}^{3N}$ is a vector describing positions, $\mathbf{q}_i \in {\mathbb R}^3$, it is shown that $U$ can be defined as a function of the interatomic distance variables $r_{ij} = |{\mathbf q}_i - {\mathbf q}_j |$, provided that the potential $U$ satisfies some symmetry assumptions. Moreover, the potential $U$ can be defined as a function of a proper subset of the distance variables~$r_{ij}$, provided that $N > 5$, with the number of distance variables used scaling linearly with the number of atoms, $N$. 
\end{abstract}

\maketitle


\section{\label{Introduction}Introduction}

\noindent
The theory of classical interatomic potentials has been developed for decades, a review of this research area is provided by Murrell et al~\cite{Murrell:1984} or more recently by Ackland~\cite{Ackland:2012}. The basis of molecular modelling is dependent on creating a suitable potential energy function that defines the free energy surface and dynamics of the system, accurately, while also balancing computational feasibility. One must compromise by reducing degrees of freedom with some method of coarse-graining~\cite{Noid:2013}. A key way to do this is by explicitly constructing a potential energy function that reduces the complexity of the system. Many such function choices can naturally arise for a given system~\cite{Hoef:1999}. Commonly pair potentials are used to approximate potential energy contributions though caution must be taken to use these appropriately~\cite{Louis:2002}. Despite this: effective pair potentials in many classical circumstances have had fair degrees of success for decades in simulations of liquids~\cite{Sprik:1993,Utterson:2022,Zhao:2019,Erban:2020,Bomont:2006,Waseda:1996}. 

To obtain more accurate results from thermodynamic calculations, many body contributions are considered in the potential energy function~\cite{Plimpton:2012,Cisneros:2016}. An example potential incorporating two-body and three-body terms is the Stillinger-Weber potential~\cite{Stillinger:1985} which accurately incorporates the geometry of silicon, meaning that not only do the pairwise bonds between the silicon atoms matter, but also the triangular sub structures connecting neighbouring atoms~\cite{Biswas:1987}. The embedded atom method potentials~\cite{Daw:1984} incorporate an effective pairwise potential and a density dependent contribution without using the geometric features explicitly. Progressing from pair potentials, to those incorporating three-body terms, and four body terms, the most general interatomic potential considered is a sum of all of these contributions, which can also include the single body terms that arise when an external field is present. The $n$-body terms are explicitly evaluated given the coordinates of the $N$ atoms: which can be thought of as vertices of a polygon (if co-planar) or a polyhedron. These $n$-body terms in the potential are then thought of as contributions arising from the $n$-gon substructures of the shape formed by the vertices. This forms the basis of fragmentation methods used in ab initio quantum chemistry, a summary and a closed form expression for energy is presented by Richard et al~\cite{Richard:2014}.
Tandem to this, cluster descriptions of many body configurations~\cite{Sanchez:1984} can also be used in conjunction with $n$-body expansions of the many-body potential~\cite{Drautz:2014}, this differs from the previous method as this relies on the ordering of vertices as opposed to their position.

Non-reciprocal interactions, where pairwise forces do not obey Newton's third law~\cite{Loos:2023}, are applicable to colloidal physics~\cite{Tsytovich:1997}, active transport~\cite{Soto:2014},  and plasma physics~\cite{Lisin:2020,Khrapak:2001}. The statistical mechanics framework used to analyse such a system that exhibits non-reciprocity relies on defining an interatomic potential: Ivlev et al~\cite{Ivlev:2015} have provided some pioneering analysis in this area.
A question that has not been answered is characterising when a general interatomic potential displays non-reciprocal interactions. For example, if the potential depends purely on pairwise distances, then reciprocity is a consequence, so one possible way to approach this problem is to study under what symmetries can we conclude that a general position dependent potential function, can be written as function that depends purely on distances. Separately, this is a fundamental question that underpins classical potential theory, and and it is addressed in this paper in Theorems~\ref{CPT} and~\ref{CPT2}. We should note that thought into symmetries of a potential has been undertaken by Kinghorn et al~\cite{Kinghorn:1997}: this was used to analyse a specific functional form of potential developed, whilst in Section~\ref{Theory} we consider a general potential with the goal of understanding when distances are appropriate variables used to describe the potential function. The potential function $U$ considered in this paper has translational, rotational and reflectional symmetries as formulated in Definition~\ref{defCPF}, where we study systems of identical particles (atoms, or more generally, coarse-grained particles). We present proofs of Theorems~\ref{CPT} and~\ref{CPT2} in Section~\ref{secproofteor}, where we also show that we only require a relatively small subset of distances to uniquely determine the potential as stated in Theorem~\ref{CPT2}. 
Limitations of this description are discussed in Section~\ref{Conclusion}, where we also present some generalizations of Theorems~\ref{CPT} and~\ref{CPT2} to mixtures of atoms of different types.

\section{Results \label{Theory}}

\noindent
The configuration of a system of $N$ atoms at positions~${\mathbf q}_i$, $i=1,2,\dots,N$, is defined as a $3N$-dimensional vector $\mathbf{q}=(\mathbf{q}_1,\mathbf{q}_2,\dots,\mathbf{q}_N)\in \mathbb{R}^{3N}$. We note that these can provisionally be thought of as vertices of an $N$-gon, or an $N$-polyhedron, assuming that ${\mathbf q}_i \ne {\mathbf q}_j$ for $i \ne j$. The lengths of edges are distances between atoms, which we denote by
\begin{equation}
r_{ij}
=
|{\mathbf q}_j - {\mathbf q}_i |,
\quad
\mbox{for}
\quad
i,j=1,2,\dots,N.
\label{distances}
\end{equation}
In this paper, we study potential functions $U:\mathbb{R}^{3N}\to \mathbb{R}$ called central potential functions which satisfy certain symmetries as specified in Definition~\ref{defCPF}. These symmetries are: (i)   
translational invariance; (ii) rotational invariance; (iii) reflectional invariance; and (iv) parity for $i$,$j$ identical atoms. An example of potential satisfying the assumptions in Definition~\ref{defCPF} is
\begin{equation}
U(\mathbf{q})=\sum\limits_{i<j}^{N}
\Psi_2(r_{ij})
+
\sum\limits_{i<j<k}^{N}
\Psi_3(r_{ij},r_{ik},r_{jk}) \, ,
\label{examplepot}
\end{equation}
where $\Psi_2 : [0,\infty) \to {\mathbb R}$ and $\Psi_3 : [0,\infty)^3 \to {\mathbb R}$ are two-body and three-body potentials which depend on distances between atoms.

\begin{widetext}
\begin{definition}
\label{defCPF}
{\rm
A function $U\!:\!\mathbb{R}^{3N}\xrightarrow{}\mathbb{R} \cup \{\pm \infty\}$ is called a {\it central potential function} provided that it takes finite values on the subset
$$
\Omega =
\left\{
\mathbf{q} \in {\mathbb R}^{3N}
\; \Big\vert \;
\mathbf{q}=\{\mathbf{q}_1,\mathbf{q}_2,...,\mathbf{q}_N\} \;
\mbox{with} \; \mathbf{q}_i \neq \mathbf{q}_j \; \mbox{for} \; i\neq j
\right\}
$$
and for any ${\mathbf q} \in \Omega$, it satisfies:
\begin{enumerate}[label=(\roman*)]
\item $U(\mathbf{q}_1+\mathbf{c},\mathbf{q}_2+\mathbf{c},\dots,\mathbf{q}_N+\mathbf{c})=U(\mathbf{q}_1,\mathbf{q}_2,\dots,\mathbf{q}_N)$ for all translations $\mathbf{c} \in \mathbb{R}^3$,
\item $U(R\,\mathbf{q}_1,R\,\mathbf{q}_2,\dots,R\,\mathbf{q}_N)=U(\mathbf{q}_1,\mathbf{q}_2,\dots,\mathbf{q}_N)$ for all rotations $R \in \mbox{SO(3)}$,
\item $U(Q\,\mathbf{q}_1,Q\,\mathbf{q}_2,\dots,Q\, \mathbf{q}_N)=U(\mathbf{q}_1,\mathbf{q}_2,\dots,\mathbf{q}_N)$ for all reflections $Q$ satisfying that all points ${\mathbf q}_i$, $i=1,2,\dots,N,$ lie on one side of the plane of reflection,
\item $U(\mathbf{q}_1,\dots,\mathbf{q}_i,\dots,\mathbf{q}_j,\dots,\mathbf{q}_N)=U(\mathbf{q}_1,\dots,\mathbf{q}_j,\dots,\mathbf{q}_i,\dots,\mathbf{q}_N)$ for any $i,j=1,2,\dots,N$.
\end{enumerate}}
\end{definition}
\end{widetext}

\noindent
The symmetries considered in Definition~\ref{defCPF} are satisfied by other generalizations of the example potential~(\ref{examplepot}), which include $n$-body terms depending only on the distances~(\ref{distances}) between atoms. In fact, the symmetries (i)-(iv) imply that the potential $U\!:\!\mathbb{R}^{3N}\to \mathbb{R}$ can be written as a function of distances. We have the following theorem which we prove in Section~\ref{secproofteor}.

\begin{theorem}
\label{CPT}
A central potential function $U:\mathbb{R}^{3N}\xrightarrow{}\mathbb{R}$ can be written as 
$$
\phi:[0,\infty)^{N(N-1)/2}\xrightarrow{}\mathbb{R},
$$ 
where the $N(N-1)/2$ inputs are interpreted as the set of all pairwise distances~$(\ref{distances})$ between atoms. 
\end{theorem}

\noindent
Considering $N=2$, Theorem~\ref{CPT} states that a central potential function $U$ of 6 variables can be written as a function $\phi$ of 1 variable, $r_{12}$. Consequently, Theorem~\ref{CPT} reduces the dimensionality of the potential $U$ for any $N<7.$ If $N=7,$ then we have $3N = N(N-1)/2 = 21$ and the $21$-dimensional state space ${\mathbb R}^{3N}$ corresponds to the 21 distance variables~(\ref{distances}). Since the dimension of the state space scales as $O(N)$ and the number of distances scales as $O(N^2)$, Theorem~\ref{CPT} can be further improved by considering only a subset of the distance variables~(\ref{distances}). In Section~\ref{secproofteor}, we also prove the following result.

\begin{theorem}
\label{CPT2}
Let $N \ge 4.$ Then a central potential function $U:\mathbb{R}^{3N}\xrightarrow{}\mathbb{R}$ can be written as 
$$
\phi: [0,\infty)^{4N-10}\xrightarrow{}\mathbb{R},
$$ 
where the $(4N-10)$ inputs are a subset of the set of all pairwise distances~$(\ref{distances})$.
\end{theorem}

\noindent
Considering $N=4$ and $N=5$, we have $4N-10=6$ and $4N-10=10$, respectively. In particular, Theorems~\ref{CPT} and~\ref{CPT2} state the same conclusion for $N=4$ and $N=5$. Theorem~\ref{CPT2} improves the result of Theorem~\ref{CPT} for $N > 5.$ We will prove Theorems~\ref{CPT} and~\ref{CPT2} together in Section~\ref{secproofteor} by considering the cases $N=2$, $N=3$, $N=4$, $N=5$ and $N > 5.$

Applying Theorem~\ref{CPT2} to our example potential~(\ref{examplepot}), we observes that it reduces the number of independent variables for $N > 5.$ In particular, while function $\phi$ constructed in the proof of Theorem~\ref{CPT2} depends only on distances~(\ref{distances}), it is not given in the form~(\ref{examplepot}).

In addition to central potential functions satisfying conditions in Definition~\ref{defCPF}, there are potentials to which Theorems~\ref{CPT} and~\ref{CPT2} are not applicable. For example, if the potential $U$ corresponds to an external non-uniform field~$\Psi_1$, then we have 
$$
U(\mathbf{q})=\sum_{i=1}^{N}
\Psi_1({\mathbf q}_i)
$$
and $U$ will neither satisfy the conditions in Definition~\ref{defCPF}, nor will it be possible to write as a function of pairwise distances~(\ref{distances}). Assuming that there is no external field present and that we have a system of $N$ identical atoms interacting (i.e. $U$ satisfies condition~(iv) in Definition~\ref{defCPF}), then we can formally write it as a sum of the $n$-body interactions for $2 \le n \le N$ in the form 
\begin{multline}
U(\mathbf{q})=\sum\limits_{i<j}^{N}U_2(\mathbf{q}_i,\mathbf{q}_j)+\sum\limits_{i<j<k}^{N}U_3(\mathbf{q}_i,\mathbf{q}_j,\mathbf{q}_k)+\dots \\
\dots+U_N(\mathbf{q}_1,\dots,\mathbf{q}_N),
\label{genswap}
\end{multline}
where we can naturally think about $n$-polyhedrons of atoms as the input to the potential function, but these are fixed in space and a natural assumption is that given this input, it should not matter where we fix this polyhedron (leading to translational invariance (i)), or how we orient this polyhedron (rotational invariance (ii)). One slightly more subtle assumption, is that we should be allowed to reflect our polyhedron in any plane that keeps the polyhedron on one side (reflectional symmetry (iii)).

It is worth noting that although these symmetries are naturally understood, they are powerful from a Hamiltonian point of view. Noether's theorem says that each symmetry gives rise to a corresponding conserved quantity (in a closed system), for example, translational invariance gives rise to conserved linear momentum (which is a consequence of reciprocity of forces). Therefore, we can intuitively understand that functions obeying these symmetries should only rely on distances. In the next section, we provide a proof of this conclusion, where we also show that a proper subset of pairwise distances for $N>5$ can be used to describe the potential function $U$. 

\renewcommand\qedsymbol{$\blacksquare$}

\section{Proofs of Theorems~\ref{CPT} and~\ref{CPT2}}
\label{secproofteor}

\noindent
We prove Theorems~\ref{CPT} and~\ref{CPT2} together by considering the cases $N=2$, $N=3$, $N=4$ and $N=5,$ followed by an inductive argument for $N>5.$ We define displacement vectors by
\begin{equation}
\boldsymbol{\Delta}_{ij}=\mathbf{q}_j-\mathbf{q}_i,
\quad
\mbox{for}
\quad
i,j=1,2,\dots,N,
\end{equation}
i.e. we have $r_{ij} = |\boldsymbol{\Delta}_{ij}|$.

Let us start with the case $N=2$. We define function $\phi: [0,\infty) \to {\mathbb R}$ by
\begin{equation}
\phi(s)
=
U(\mathbf{0},s\hat{\textbf{k}})
=
U(0,0,0,0,0,s),
\label{phiN2}
\end{equation}
where $\hat{\textbf{k}}$ is a unit vector in the direction of the positive $z$ axis and $\mathbf{0}=[0,0,0].$ Given atom positions $\mathbf{q}_1,\mathbf{q}_2  \in \mathbb{R}^3$, we translate the configuration to position atom 1 at the origin. Using  symmetry (i) in Definition~\ref{defCPF}, we have $U(\mathbf{q}_1,\mathbf{q}_2)=U(\mathbf{0},\boldsymbol{\Delta}_{12})$. We then rotate the axes using rotation $R_1 \in \mbox{SO}(3)$ such that the displacement vector connecting the two atoms is aligned with the positive $z$ axis, giving $R_1 \boldsymbol{\Delta}_{12}=r_{12}\hat{\textbf{k}}$, while maintaining $R_1\mathbf{0}=\mathbf{0}$. Using symmetry (ii) in Definition~\ref{defCPF}, we have
$$
U(\mathbf{q}_1,\mathbf{q}_2)=U(\mathbf{0},\boldsymbol{\Delta}_{12})
=
U(\mathbf{0},r_{12}\hat{\textbf{k}})
=
\phi(r_{12}),
$$
where the last equality follows from our definition~(\ref{phiN2}). This concludes the proof of Theorem~\ref{CPT} for $N=2.$
 
\subsection{The case $N=3$}

\label{secN3}

\begin{figure}[!]%
\includegraphics[width=0.47\textwidth]{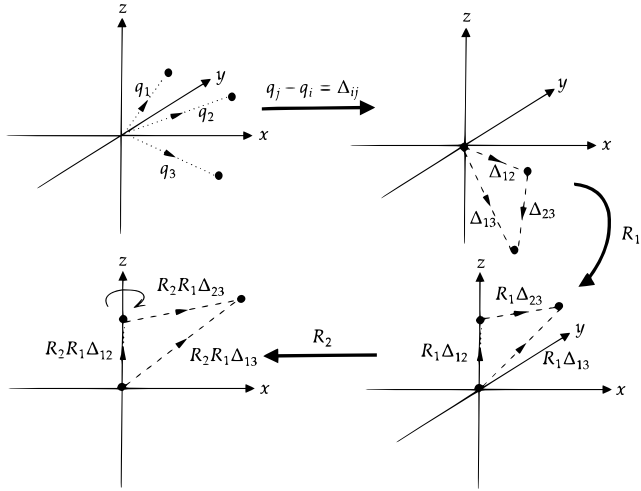}\;
\caption{{\it A schematic of the constructive method in aid of the proof for the case $N=3$.}\label{N3caseproof}}
\end{figure}

\noindent
Given atom positions $\mathbf{q}_1,\mathbf{q}_2,\mathbf{q}_3  \in \mathbb{R}^3$, we consider the function $U(\mathbf{q}_1,\mathbf{q}_2,\mathbf{q}_3)$. Using symmetry (i) in Definition~\ref{defCPF}, we translate the configuration to position atom 1 at the origin and consequently, we have 
$$
U(\mathbf{q}_1,\mathbf{q}_2,\mathbf{q}_3)
=
U(\mathbf{0},\boldsymbol{\Delta}_{12},\boldsymbol{\Delta}_{13}).
$$
Given that we have three axes to rotate around, we can always find a rotation $R_1$ such that $R_1\Delta_{12}=r_{12}\hat{\textbf{k}}$, as we did in the $N=2$ case. Using symmetry (ii), we have $$
U(\mathbf{0},\boldsymbol{\Delta}_{12},\boldsymbol{\Delta}_{13})
=
U(\mathbf{0},r_{12}\hat{\textbf{k}},R_1\boldsymbol{\Delta}_{13}).
$$
We note that we can find a rotation $R_2$ about the $z$ axis which rotates the triangle defined by the transformed atom positions $\mathbf{0}$, $r_{12}\hat{\textbf{k}}$ and $R_1\boldsymbol{\Delta}_{13}$ to a planar triangle in the $x$-$z$ plane, as demonstrated in Figure~\ref{N3caseproof}. Using symmetry (ii) again, we have
$$
U(\mathbf{q}_1,\mathbf{q}_2,\mathbf{q}_3)
=
U(\mathbf{0},
r_{12}\hat{\textbf{k}},
R_2 R_1 \boldsymbol{\Delta}_{13}).
$$
However the key point is that  $R_2 R_1 \boldsymbol{\Delta}_{13}$ is uniquely defined by the triangle with lengths $r_{12}$, $r_{13}$ and $r_{23}$, the angles of which can be calculated using the cosine rule, i.e. 
$
R_2 R_1 \boldsymbol{\Delta}_{13}
$
can be expressed as
\begin{equation}
\left[ 
\sqrt{r_{13}^2-
\left(\frac{r_{13}^2 + r_{12}^2 - r_{23}^2}{2 r_{12}}
\right)^2}, \,
0 \, ,
\frac{r_{13}^2 + r_{12}^2 - r_{23}^2}{2 r_{12}} 
\right].
\label{r21r31r32}
\end{equation}
Therefore there exists function $\phi: [0, \infty)^3 \to {\mathbb R}$ such that
$U(\mathbf{q}_1,\mathbf{q}_2,\mathbf{q}_3)=\phi(r_{12},r_{13},r_{23}),$ for any $\mathbf{q}_1,$ $\mathbf{q}_2$ and $\mathbf{q}_3$, confirming Theorem~\ref{CPT} for $N=3$.

\subsection{The case $N=4$}

\label{secN4}

\begin{figure}[!]
\includegraphics[width=0.47\textwidth]{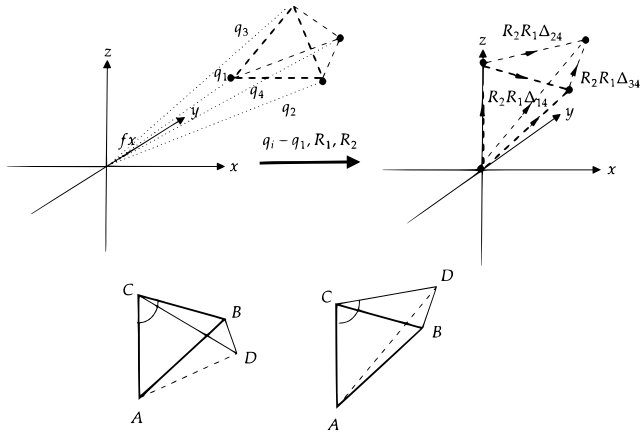}\;
\caption{{\it A schematic of the constructive method in aid of the proof for $N=4$. For clarity we have only highlighted the additional three displacement vectors, though the triangle formed by vertices $\{1,2,3\}$ lying in the $x$-$z$ plane is the same as in Figure~\ref{N3caseproof}.}\label{N4caseproof}}
\end{figure}

\noindent
Given atom positions $\mathbf{q}_1,\mathbf{q}_2,\mathbf{q}_3,\mathbf{q}_4  \in \mathbb{R}^3$, these can be thought of defining the vertices of a tetrahedron (or if co-planar a quadrilateral). Following similar steps as in the case $N=3$ in Section~\ref{secN3}, we translate atom 1 to the origin, apply rotation $R_1$ to orient displacement vector $\boldsymbol{\Delta}_{12}$ with the positive $z$ axis, then do a second rotation $R_2$ that fixes the triangle formed by the vertices of atoms 1, 2 and 3 in the $x$-$z$ plane. As in Section~\ref{secN3}, we have 
$$
U(\mathbf{q}_1,\mathbf{q}_2,\mathbf{q}_3,\mathbf{q}_4)
=
U(\mathbf{0},
r_{12}\hat{\textbf{k}},
R_2 R_1 \boldsymbol{\Delta}_{13},
R_2 R_1 \boldsymbol{\Delta}_{14}).
$$
Using equation~(\ref{r21r31r32}), we know that $R_2 R_1 \boldsymbol{\Delta}_{13}$ is determined entirely by distances $r_{12},$ $r_{13}$ and $r_{23}$. All that remains to be shown is that $R_2 R_1 \boldsymbol{\Delta}_{14}$ is determined by pairwise distances. We note that the triangle formed by atoms 1, 2 and 3 (denoted as $ABC$ in the lower part of our illustration of the proof in Figure~\ref{N4caseproof}) is uniquely determined (after orienting one side with the positive $z$ axis). Consequently, this fixes the side $BC$. On the other hand the triangle $BCD$ is uniquely determined (as one side $BC$ is fixed) by distances $r_{23}$, $r_{24}$ and $r_{34}$. 
These can be thought of as two triangles which can rotate around a hinge $BC$, so to determine the vector $R_2 R_1 \boldsymbol{\Delta}_{14}$, we necessarily need the final distance $r_{14}$ that gives the angle between the planes containing triangles $ABC$ and $BCD$ (two configurations are illustrated in Figure~\ref{N4caseproof}). If triangles $ABC$ and $BCD$ are co-planar, the set of all pairwise distances, with this orientation, will give a unique description of $R_2 R_1 \boldsymbol{\Delta}_{14}$. If these triangles are not co-planar, this final distance gives two possible vectors for $R_2 R_1 \boldsymbol{\Delta}_{14}$. These correspond to a unique $R_2 R_1 \boldsymbol{\Delta}_{14}$ and the copy obtained by reflection in the plane containing triangle $ABC$. However by property (iii) we know that if we reflect in the plane containing $ABC$ with a matrix denoted $Q$, then 
\begin{eqnarray*}
&&    
U(\mathbf{0},
r_{12}\hat{\textbf{k}},
R_2 R_1 \boldsymbol{\Delta}_{13},
R_2 R_1 \boldsymbol{\Delta}_{14})
\\
&&
\qquad =
U(\mathbf{0},
r_{12}\hat{\textbf{k}},
R_2 R_1 \boldsymbol{\Delta}_{13},
Q R_2 R_1 \boldsymbol{\Delta}_{14})
\end{eqnarray*}
Therefore there exists function $\phi: [0, \infty)^6 \to {\mathbb R}$ such that
$
U(\mathbf{q}_1,\mathbf{q}_2,\mathbf{q}_3,\mathbf{q}_4)
=
\phi(r_{12},r_{13},r_{14},r_{23},r_{24},r_{34}),$ for any $\mathbf{q}_1,$ $\mathbf{q}_2$, $\mathbf{q}_3$ and $\mathbf{q}_4$, confirming Theorem~\ref{CPT} for $N=4$.

\subsection{The case $N=5$}

\label{secN5}

\noindent
To proceed in this case, we note that any $N$ vertex polyhedron can be made by adding a single vertex to an $N-1$ polyhedron or polygon (in the case where all other points are co-planar). The task at hand, as in the case $N=4$ in Section~\ref{secN4}, is being able to determine the displacement vectors once we have translated and rotated the configuration such that $R_2 R_1 \boldsymbol{\Delta}_{12} = r_{12}\hat{\textbf{k}}$ is aligned with the positive $z$ axis. 

\begin{figure*}[t]%
(a) \hskip 9cm (b) \hfill\break
\centerline{
\includegraphics[width=0.495\textwidth]{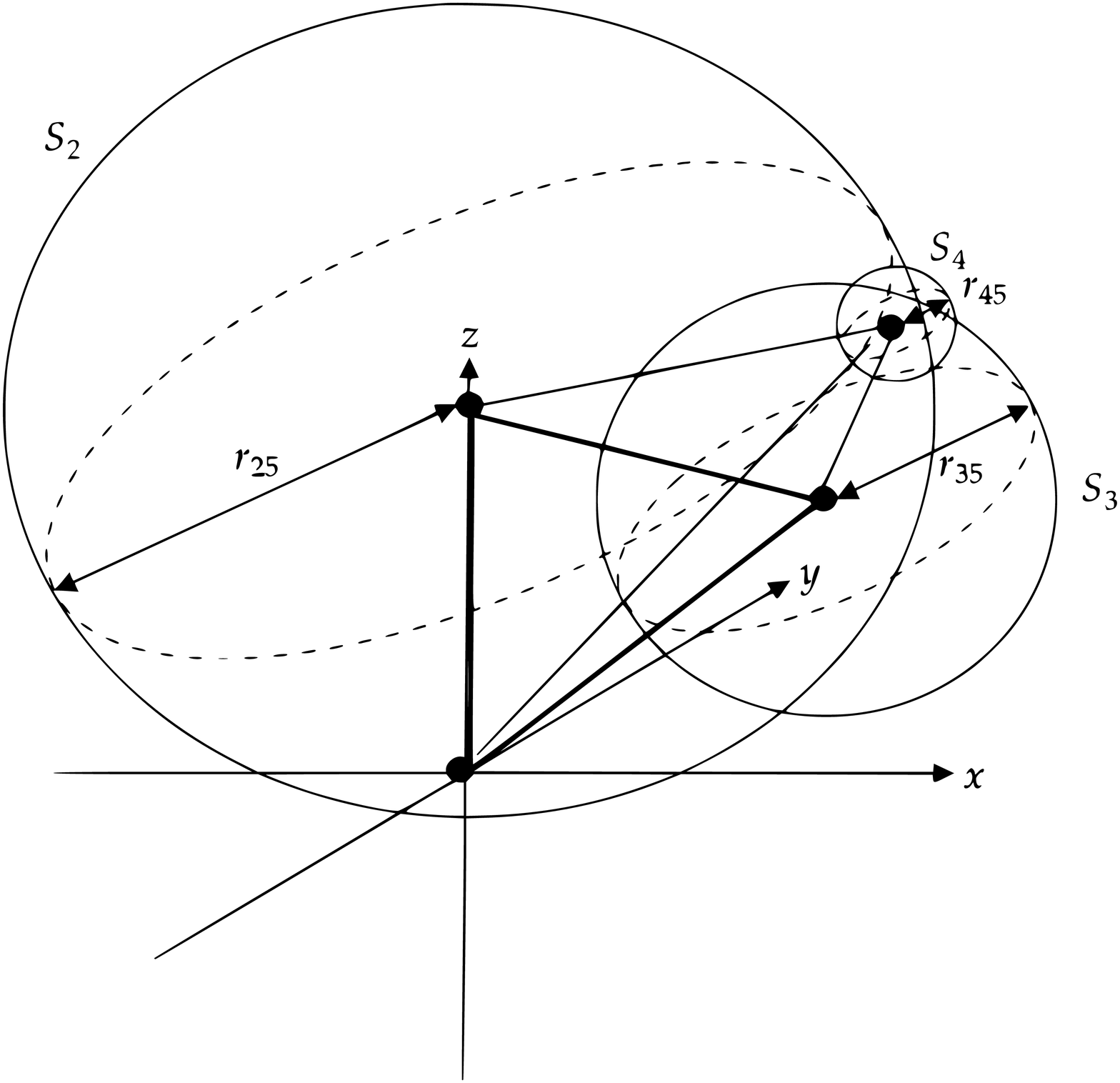}\;
\includegraphics[width=0.495\textwidth]{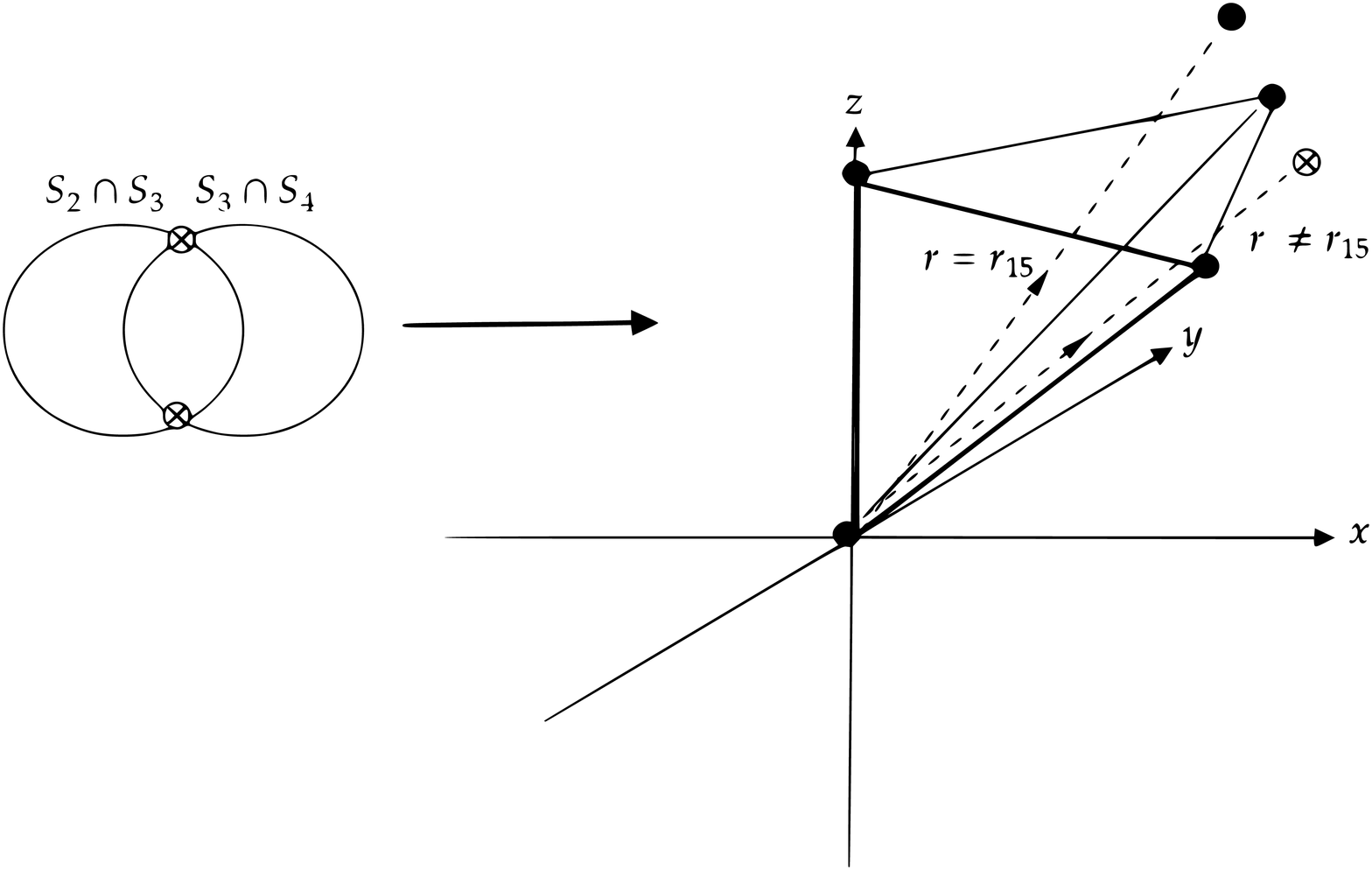}}
\caption{{\rm (a)} {\it The intersection of $3$ spheres, based on three known centres and radii are used to position an additional vertex.}\hfill\break {\rm (b)} {\it Here a fourth vertex, chosen non co-planar to the three vertices used to construct the spheres previously, is used to uniquely determine the fourth vertex position.}
}\label{N5caseproof}
\end{figure*}

An $N=5$ polyhedron can be constructed from either adding a vertex onto a pre-existing $N=4$ polyhedron (at most 3 points are co-planar), or an $N=4$ polygon (where all points are co-planar). In the first case, we may take any 3 vertices on the pre-existing polyhedron: call these vertices the transformed positions of atoms $2,3,4$ (by property (iv) in Definition~\ref{defCPF}). If we know $r_{25}$, then this fifth vertex must lie on a sphere of radius $r_{25}$, with the transformed position of atom 2 as the centre: we denote this $S_2$. Similarly, we construct $S_3$ and $S_4$ as spheres of radii $r_{35}$ and $r_{45}$ respectively. This is illustrated in Figure~\ref{N5caseproof}(a). 
The fifth vertex lies at the intersection of three spheres $S_2$, $S_3$ and $S_4$, which contains at most two points. If it contains exactly two points, then we need another distance $r_{15}$ (which is the distance from the vertex in the pre-existing polyhedron that was not used as a centre of spheres $S_2$, $S_3$ or $S_4$) to determine which of those two positions is correct, see Figure~\ref{N5caseproof}(b). In this way: 4 more distances are used to specify all of the vertices of the $N=5$ polyhedron. Therefore, there exists function $\phi: [0, \infty)^{10} \to {\mathbb R}$ such that
$
U(\mathbf{q}_1,\mathbf{q}_2,\mathbf{q}_3,\mathbf{q}_4,{\mathbf q}_5)
=
\phi(r_{12},r_{13},r_{14},r_{15},r_{23},r_{24},r_{25},r_{34},r_{35},r_{45}),$ for any $\mathbf{q}_1,$ $\mathbf{q}_2$, $\mathbf{q}_3$, ${\mathbf q}_4$ and $\mathbf{q}_5$. 

To arrive at this conclusion, we used an assumption that no four points are co-planar. If this is not the case, then we need less distances for the specific configuration. For example, if the pre-existing 4 vertices are co-planar: utilising the sphere approach for any three of those vertices will result again in two possible positions for vertex 5, however using the pairwise distance between this vertex and the new vertex gives no information, as the fourth point lies on the plane of symmetry formed by the spheres. In this case we use  property (iii), considering the reflective symmetry about this plane to argue that we have determined all displacement vectors with this orientation uniquely up to a reflection in the plane containing vertices 1, 2, 3 and 4. In this case, we do not need the fourth distance mentioned above, and evaluating $U(\mathbf{q}_1,\mathbf{q}_2,\mathbf{q}_3,\mathbf{q}_4,\mathbf{q}_5)$ is possible with the 9 pairwise distances. The tenth distance is also not needed if the intersection of spheres $S_2,$ $S_3$ and $S_4$ is exactly equal to one point (when vertices 2, 3 and 4 are co-linear). Thus we have proven Theorems~\ref{CPT} and~\ref{CPT2} in the case $N=5$.

\subsection{The case $N>5$}

\noindent
We inductively prove that a similar setup as in the $N=5$ case in Section~\ref{secN5} works by constructing polyhedra of higher order by the addition of a new vertex. Say that the $N-1$ case required the set of distances $D_{N-1}$ to evaluate $U(\mathbf{q}_1,\dots,\mathbf{q}_{N-1})$, where $|D_{N-1}|= 4(N-1)-10$. Which is true for the base case of $N=5$.

The most general case to consider is when we have an $N-1$ polyhedron before we introduce the new vertex. In this case, any three of the $N-1$ vertices can be chosen, say $i,j,k$. The three distances $r_{iN},r_{jN},r_{kN}$ are used to create three intersecting spheres and two potential positions for vertex $N$. We use vertex $l$, which is not co-planar to $i,j,k$ 
and distance $r_{lN}$ determines this position uniquely. Therefore the required set of distances to evaluate $U(\mathbf{q}_1,\dots,\mathbf{q}_{N})$ is $D_N=D_{N-1} \cup\{r_{iN},r_{jN},r_{kN},r_{lN}\}$, i.e. we need $4N-10$ pairwise distances. If there are at least four co-planar points, then we only need 3 additional pairwise distances (so we would only need $4N-11$ pairwise distances). Since the inductive step holds for all $N$, and it works for the base case of $N=5$, this concludes our proofs of Theorems~\ref{CPT} and~\ref{CPT2} for all $N$.

\section{Discussion and Conclusions} 
\label{Conclusion}

\noindent
Theorems~\ref{CPT} and~\ref{CPT2} show that symmetries of the many-body system imply that the potential $U$ can be written in a form which only depends on pairwise distances between atoms. Since $U$ is a function of the $3N$-dimensional state space, Theorem~\ref{CPT} provides a non-linear transformation of $U$ from a function of $3N$ variables ${\mathbf q}$ into a function of $N(N-1)/2$ distance variables, which is not optimal as it is shown in Theorem~\ref{CPT2} where the number of distance variables scales linearly with $N.$

Considering the example potential~(\ref{examplepot}), which depends on all $N(N-1)/2$ distance variables, Theorem~\ref{CPT2} provides a reduction of the number of distance variables to $O(N).$ However, if we use the resulting form of the potential, $\phi$, this does not directly translate to $O(N)$ complexity of evaluating $\phi$. To illustrate this, let us consider our example potential~(\ref{examplepot}) with $\Psi_2(r_{ij})=1/r_{ij}$ and $\Psi_3 \equiv 0,$ giving
\begin{equation}
U(\mathbf{q}_1,\mathbf{q}_2,\dots,\mathbf{q}_N)
=
\sum\limits_{\{i,j\} \in {\cal S}} 
\frac{1}{r_{ij}}
+
\sum\limits_{\{i,j\} \not \in {\cal S}} 
\frac{1}{r_{ij}},
\label{limitSnotS}
\end{equation}
where $r_{ij}$ is defined by~(\ref{distances}) and ${\cal S}$ is the set of pairs of indices $\{i,j\}$ corresponding to the subset of distances which is used to define $\phi$ in Theorem~\ref{CPT2}. 

Theorem~\ref{CPT2} shows that the number of elements in the set ${\cal S}$ scales as $O(N)$, i.e. the number of terms in the first sum on the right hand side of~(\ref{limitSnotS}) is $O(N)$, while the number of terms in the second sum on the right hand side of~(\ref{limitSnotS}) scales as $O(N^2)$. Considering $\{i,j\} \not \in {\cal S}$, we can find $k \in \{ 1, 2, \dots, N\}$ such that $\{i,k\} \in {\cal S}$ and $\{j,k\} \in {\cal S}$. In particular, distance $r_{ij}$ for $\{i,j\} \not \in {\cal S}$ can be expressed in terms of distances $r_{ik}$ and $r_{jk}$ using the cosine rule. Therefore, we can find an explicit form of the potential~$U$ as a function of $O(N)$ distances corresponding to the indices in the set ${\cal S}$. However, the second term in the form~(\ref{limitSnotS}) will contain summations over $O(N^2)$ terms. That is, Theorem~\ref{CPT2} does not reduce the $O(N^2)$ complexity of calculations of $\phi$. It has been included to illustrate that the number of distance variables needed scales linearly with $N$ in the same way as the dimension of the phase space scales linearly with $N.$

Theorem~\ref{CPT} has been formulated as an implication, stating that symmetries (i)-(iv) of a central potential function in Definition~\ref{defCPF} imply that the potential can be written as a function of pairwise distances~(\ref{distances}). However, translations, rotations and reflections are Euclidean isometries, preserving pairwise distances between atoms, so a partial inverse of Theorem~\ref{CPT} also holds, i.e. any potential given as a function of pairwise distances satisfies symmetry assumptions~(i)-(iii). The property (iv) states that we consider systems of identical particles in this paper. In particular, symmetries (i)-(iv) are both necessary and sufficient conditions for a potential to be expressed as a function of pairwise distances for systems of identical atoms. 

We can generalize Theorems~\ref{CPT} and~\ref{CPT2} to mixtures of particles, i.e. for systems when symmetry (iv) in Definition~\ref{defCPF} does not hold. Then properties (i)-(iii) of potential~$U$ imply that it can be expressed as a function of pairwise distances. If we further reduce the number of symmetries the potential $U$ has, then we can find potential functions which cannot be expressed as a function of pairwise distances. For example, 
$$
U(\mathbf q) 
= 
U({\mathbf q}_1, {\mathbf q}_2, \dots, {\mathbf q}_N)
=
{\mathbf d} \cdot ({\mathbf q}_2 - {\mathbf q}_1)
$$ for any nonzero constant vector ${\mathbf d}$ satisfies the translational symmetry~(i), but not the rotational symmetry~(ii). An example of potential function satisfying the rotational symmetry (ii) but not the translational symmetry (i) is $U(\mathbf q) = | \mathbf{q_1} |.$ In fact, symmetries (i)-(iii) are both necessary and sufficient conditions for a potential to be expressed as a function of pairwise distances (for systems of non-identical particles).

\vskip 2mm

\rule{0pt}{1pt}

\begin{acknowledgments}
\noindent
This work was supported by the Engineering and Physical Sciences Research Council (EPSRC) [grant number EP/V047469/1] and by the Royal Society [grant number RGF\textbackslash EA\textbackslash 180058]. 
\end{acknowledgments}


\begin{thebibliography}{10}

\bibitem{Murrell:1984}
J.~Murrell et al.
\textit{Molecular potential energy functions}.
Wiley (1984)

\bibitem{Ackland:2012}
G.~Ackland.
\textit{Interatomic potential development}.
Comprehensive Nuclear Materials, pp. 267-291 (2012)

\bibitem{Noid:2013}
W.~Noid.
\textit{Perspective: coarse-grained models for bio\-molecular systems}.
Journal of Chemical Physics {\bf 139}(9), 090901 (2013)

\bibitem{Hoef:1999}
M.~Hoef and P.~Madden.
\textit{Three-body dispersion contributions to the thermodynamic properties and effective pair interactions in liquid argon}.
Journal of Chemical Physics {\bf 111}(4), pp. 1520-1526 (1999) 

\bibitem{Louis:2002}
A~Louis.
\textit{Beware of density dependent pair potentials}.
Journal of Physics: Condensed Matter {\bf 14}, pp. 9187 (2002)

\bibitem{Sprik:1993}
M.~Sprik
\textit{Effective pair potentials and beyond}. In: Computer Simulation in Chemical Physics. 
NATO ASI Series, vol 397, Springer, Dordrecht (1993)

\bibitem{Zhao:2019}
C.~Zhao et al.
\textit{Seven-site effective pair potential for simulating liquid water}.
Journal of Physical Chemistry B {\bf 123}(21), pp. 4594–4603 (2019)

\bibitem{Bomont:2006}
J.~Bomont and J.~Bretonnet.
\textit{An effective pair potential for thermodynamics and structural properties of liquid mercury}.
Journal of Chemical Physics {\bf 124}, 054504 (2006)

\bibitem{Utterson:2022}
J. Utterson and R. Erban.
\textit{On standardised moments of force distribution in simple liquids}. 
Physical Chemistry Chemical Physics {\bf 24}, pp. 5646-5657 (2022) 

\bibitem{Waseda:1996}
K.O. Waseda.
\textit{Structure and effective pair potential of liquid silicon}.
Japanese Journal of Applied Physics {\bf 35}, pp. 151 (1996)

\bibitem{Erban:2020}
R.~Erban and S.J.~Chapman.
\textit{Stochastic modelling of reaction-diffusion processes}. Cambridge Texts in Applied Mathematics, Cambridge University Press (2020) 

\bibitem{Plimpton:2012}
S.~Plimpton and A.~Thompson. 
\textit{Computational aspects of many-body potentials}.
MRS Bulletin. Cambridge University Press {\bf 37}(5), pp. 513–521 (2012)

\bibitem{Cisneros:2016}
G.~Cisneros et al.
\textit{Modeling molecular interactions in water: from pairwise to many-body potential energy functions}.
Chemical Reviews {\bf 116}(13), pp. 7501–7528 (2016)

\bibitem{Stillinger:1985}
F.~Stillinger and T.~Weber.
\textit{Computer simulation of local order in condensed phases of silicon}.
Physical Review B {\bf 31}(8), pp. 5262 (1985)

\bibitem{Biswas:1987}
R.~Biswas and D.~Hamann.
\textit{New classical models for silicon structural energies}.
Physical Review B {\bf 36}(12), pp. 6434 (1987)

\bibitem{Daw:1984}
M.~Daw and M.~Baskes.
\textit{Embedded-atom method: Derivation and application to impurities, surfaces, and other defects in metals}.
Physical Review B {\bf 29}(12), pp. 6443 (1984)

\bibitem{Richard:2014}
R.~Richard, K.~Lao and J.~Herbert.
\textit{Understanding the many-body expansion for large systems. I. precision considerations}.
Journal of Chemical Physics {\bf 141}(1), 014108 (2014)

\bibitem{Sanchez:1984}
J.~Sanchez, F.~Ducastelle and D.~Gratias.
\textit{Generalized cluster description of multicomponent systems}.
Physica A: Statistical Mechanics and its Applications {\bf 128}(1-2), pp. 334-350 (1984)

\bibitem{Drautz:2014}
R.~Drautz, M.~F\"ahnle, and J.~Sanchez.
\textit{General relations between many-body potentials and cluster expansions in multicomponent systems}.
 Journal of Physics: Condensed Matter {\bf 16}(23), pp. 3843 (2004)

\bibitem{Loos:2023}
S.~Loos et al.
\textit{Nonreciprocal forces enable cold-to-hot heat transfer between nanoparticles}.
Scientific Reports {\bf 13}, 4517 (2023)

\bibitem{Tsytovich:1997}
V.~Tsytovich.
\textit{Dust plasma crystals, drops, and clouds}.
Physics-Uspekhi {\bf 40}(1), pp. 53 (1997)

\bibitem{Lisin:2020}
E.~Lisin, O.~Petrov, and E.~Sametov.
\textit{Experimental study of the nonreciprocal effective interactions between microparticles in an anisotropic plasma}. 
Scientific Reports {\bf 10}, 13653 (2020)

\bibitem{Khrapak:2001}
S.~Khrapak, A.~Ivlev and G.~Morfill.
\textit{Interaction potential of microparticles in a plasma: role of collisions with plasma particles}.
Physical Review E {\bf 64}(4), 046403 (2001)

\bibitem{Soto:2014}
R.~Soto and R.~Golestanian.
\textit{Self-assembly of catalytically active colloidal molecules: tailoring activity through
surface chemistry}.
Physical Review Letters {\bf 112}(6), 068301 (2014)

\bibitem{Ivlev:2015}
A.~Ivlev et al.
\textit{Statistical mechanics where Newton’s third law is broken}.
Physical Review X {\bf 5}, 011035 (2015)

\bibitem{Kinghorn:1997}
D.~Kinghorn and L.~Adamowicz.
\textit{A new N-body potential and basis set for adiabatic and non-adiabatic variational energy calculations}. Journal of Chemical Physics {\bf 106}(21), pp. 8760-8768 (1997) 

\end{thebibliography}
\end{document}